\documentclass[aps,12pt,superscriptaddress,prl,floatfix]{revtex4-2}

\usepackage{bm}
\usepackage{graphicx}
\usepackage{subfig}
\usepackage{natbib}
\usepackage{siunitx}
\usepackage{mathtools}
\usepackage{bbold} 
\usepackage{color}
\def\be{\begin{equation}}
\def\ee{\end{equation}}
\def\e#1{\label{#1}\end{equation}}
\def\bea{\begin{eqnarray}}
\def\eea{\end{eqnarray}}
\def\ea#1{\label{#1}\end{eqnarray}}

\def\bem#1{\begin{mathletters}\label{#1}}
\def\eml{\end{mathletters}}

\newcommand{\ket}[1]{\left| #1 \right>}

\def\braket#1#2{{\langle#1|#2\rangle}}

\def\4#1{{\boldsymbol{#1}}}
\def\8#1{{\widetilde{#1}}}
\def\bse{\begin{subequations}}
\def\ese{\end{subequations}}

\newcommand{\mathsym}[1]{{}}
\newcommand{\unicode}[1]{{}}

\begin{document}

\title{Frequency measurements beyond the Heisenberg time-energy limit with a single atom}

\author{Liam P. McGuinness}
\email{liam.mcguinness@anu.edu.au}
\affiliation{Laser Physics Centre, Research School of Physics, Australian National University, Acton, Australian Capital Territory 2601, Australia}

\begin{abstract}
The Heisenberg time-energy relation prevents determination of an atomic transition to better than the inverse of the measurement time. The relation generally applies to frequency estimation of a near-resonant field [1-3], since information on the field frequency can be used to infer the atomic transition [4, 5]. Here we demonstrate a frequency estimation technique that provides an uncertainty orders of magnitude below the Heisenberg limit with a single atom. With access to $N$ atoms, we propose a fundamental uncertainty limit improving as $\sqrt{N}$, regardless of whether entanglement is employed. We describe implementation of the quantum fourier transform to estimate an unknown frequency without using entanglement. A comparison to classical algorithms severely limits the benefit that quantum algorithms provide for frequency estimation and that entanglement provides to quantum sensing in general.
\end{abstract}

\maketitle

\newpage
The time-energy (T-E) uncertainty relation fundamentally limits how precisely one can estimate an unknown atomic transition $E$, in a given time $T$. As a consequence, physical devices that use a measurement of $E$ to infer another parameter, have an uncertainty governed by this principle. Noting different interpretations [6, 7], if we accept $\Delta E\, T \geq \hbar$, where $\Delta E$ is the uncertainty in $E$ and $\hbar$ is the reduced Planck constant, one can associate $\omega_0 = E/\hbar$, to obtain a time-frequency relation (connected to the Gabor-Fourier limit [8, 9]) which prevents estimation of the atomic transition frequency $\omega_0$ to better than 1\,Hz in a second [1]. Equipped instead with an atom of known energy, measurement of the detuning from resonance allows the frequency $\omega$ of a near-resonant field to be estimated with the same precision.

Here we show that the apparent symmetry between these two scenarios is not exact. In doing so, we use a single atom to estimate $\omega$ with an uncertainty beyond the T-E limit. The violation is achieved by \emph{i)} identifying and estimating a parameter independent of $E$, for which the T-E relation does not hold and \emph{ii)} use of multiple measurements during $T$ to improve the overall precision beyond that of any individual estimate. That \emph{ii)} is required, follows if no individual measurement provides an uncertainty better than $1/T$, which we refer to as the 1TF conjecture as we are unaware of a proof. As the 1TF conjecture is crucial to the following analysis and likely to be a point of debate, further discussion is included in SI Note 1. The two criteria can be immediately applied to rule out proposals with super-linear or exponential uncertainty reduction for surpassing the T-E limit. Using quantum mechanical and information theoretic principles we bound the uncertainty of any frequency estimation algorithm implemented on a physical system.

First we describe how measurement of the signal frequency $\omega$, can be decoupled from $\omega_0$. The key is to estimate the signal phase using an estimator that is insensitive to $E$. To see this, we write the Hamiltonian describing the atomic spin and interaction with a near-resonant field from $t_0$ to $t$:
\begin{equation}
H(t_0,t)=\frac{\hbar \omega_0}{2}\sigma_z + \frac{\hbar \Omega}{2} \left( \cos \left[ \omega t + \varphi (t_0) \right] \sigma_x + \sin \left[ \omega t + \varphi (t_0) \right] \sigma_y \right),
\end{equation}
where $\Omega$ is the Rabi frequency of the atom-field interaction, $\varphi(t_0)$ the initial phase of the field, and $\sigma_{x,y,z}$ the Pauli spin-matrices. Where required $s,c$ subscripts are used to distinguish signal and control fields respectively, and we define { }\(\omega _{s,c} =\omega _0+\delta_{s,c}\) as their detuning from atomic resonance, with a relative detuning \(\delta _s = \delta _c+\delta\). We set $\Omega_s= \Omega_c \equiv \Omega$ (taken to be real, positive and known \emph{a priori}), and $\varphi_c (t_0 )=0$. If $\delta_{s,c} < \Omega$ the field is near-resonant. As only the signal phase depends on $t_0$, for concision we write $\varphi$ with the understanding that it depends on the starting time of the experiment.

For an atom initially in state $\ket{\psi_0} = \ket{0}$ (where $\ket{0}$, $\ket{1}$ denote the eigenstates along $z$), interaction with the signal from $t_0=0$ to $t=\pi/(2\Omega)$, performs a one-to-one mapping of $\varphi$ to a state $\ket{\psi} \approx \frac{1}{\sqrt{2}} \left( \ket{0} + e^{i(\varphi-\pi/2)} \ket{1}\right)$ in the $x-y$ plane, defined in a frame rotating with the signal frequency (Fig. 1a). Although exact only for $\delta_s=0$, the mapping error is independent of $\delta_s$ to first order, and thus the spin transition energy (see methods). It is this fact that we exploit here. Importantly and counterintuitively, insensitivity to $\delta_s$, and thus $\omega_0$, does not reduce the estimation accuracy of the field frequency. Readout and estimation of signal phase can then be performed by measuring the spin population along $x$. Experimentally this is done by using a second $\pi/2$-rotation mediated by the control field, and population readout along $z$, to obtain $\hat{\varphi} = \cos^{-1}  \left[ 2 \mathrm{Pr}(1)-1 \right]$, where Pr$(1)=1 - \left|\braket{\psi}{\psi_0}\right|^2 \approx \cos^2\left[\varphi/2 \right]$ is the probability to measure the atom in $\ket{1}$ (see SI Note 2). Here, and in contrast to atomic clocks, the second $\pi/2$-rotation is performed immediately after the initial $\pi/2$-rotation, and mediated by a control field with known frequency (Fig. 2a).

Experiments are performed on a single nitrogen-vacancy (NV) center in diamond and the signal phase is mapped to the qubit state by resonantly driving the NV spin with a microwave field of $\omega \sim 1.5$\,GHz. The negatively charged NV center has an electronic spin-1 ground state, with coherence times up to a millisecond at room-temperature [10, 11]. Optical readout of the NV spin state is performed with a confocal microscope by applying a 350\,ns green laser pulse and recording the spin-dependent photoluminescence intensity (Fig. 1b). Near a magnetic field of 500\,G, the transition frequency between the $\ket{0}, \ket{1}$ spin-states coincides with the microwave frequency. At this field the nuclear spin of the NV center is polarized [12], and off-resonant transitions are more than 1\,GHz away, allowing for treatment as a two-level qubit. A Rabi frequency of 40\,MHz, corresponding to a 6\,ns $\pi/2$-rotation is obtained by using a coplanar stripline in contact with the diamond.

In Fig. 2b, the phase estimate, $\hat{\varphi}$ is shown for a signal with fixed initial phase and frequency ($\varphi =0.95, \omega_s= 1.511406$\,GHz), where the external magnetic field was shifted up to 46\,G to detune the spin transition frequency from resonance. A control field, co-resonant with the signal was used to readout the spin population. For magnetic field shifts up to 1\,G, corresponding to a signal detuning from atomic resonance of several MHz, the estimated phase is insensitive to the magnetic field. In addition to static shifts, $\hat{\varphi}$ is insensitive to magnetic fluctuations or even spin relaxation at MHz rates. This estimator independence of sensor coherence time and intrinsic noise resilience is not shared by other quantum metrology techniques. From error analysis (SI Note 3), we find that the statistical error in $\hat{\varphi}$ depends on fourth power of the qubit detuning (Fig. 2b middle panel). For shot noise of magnitude $\sigma_{SN}= 1/8$, the statistical error of a single measurement averaged over $\varphi \in [0,2\pi]$ is $\Delta \hat{\varphi}_{stat} \approx 1+\frac{(\pi-4)^2\delta _s{}^4}{16 \Omega ^4} + O\left[\delta_s \right]^5$ where the same error is obtained for quantum projection noise $\sigma_{QPN} = \sqrt{\mathrm{Pr}(1)(1-\mathrm{Pr}(1))}$. The systematic error is $\Delta \hat{\varphi}_{sys} \approx  \frac{2 \varphi  \delta _s}{\Omega }+O\left[\delta _s\right]{}^2$, i.e. approximately a factor of $\Omega$ less than the qubit detuning (Fig. 2b, lower panel).

Whilst Fig. 2b demonstrates estimation of the signal phase with a uncertainty that is decoupled from $E$, typically in quantum sensing, the atomic frequency is known and held constant to within the decoherence time and the control tuned to this frequency. In the case $\delta_c=0$, for a signal with detuning $\delta_s< \Omega$, the statistical and systematic errors in $\hat{\varphi}$ are the same as obtained above (SI Note 3). Fig. 2c shows the atomic response to the signal phase as a function of $\delta_s$, obtained by varying $\varphi$ from $0\,–\,2\pi$, and plotting the change in atomic population (Fig. 2c lower). We note two features of the response curve, namely it is quadratically dependent on the signal frequency, and maximal sensitivity to $\varphi$ is obtained when $\delta_s=0$ (see SI Note 3). These observations justify treating the signal as perfectly resonant to the atom during interaction in deriving frequency estimation bounds, with the knowledge that the bounds cannot be reached for signals with finite detuning, but that they can be approached due to the weak dependence on $\delta_s$.

How well can one employ knowledge of $\varphi$ to estimate $\omega$? From $\hat{\omega}=\left(\hat{\varphi}(t_1) - \hat{\varphi}(t_0)\right)/(t_1-t_0)$, if $\varphi(t_0)$ is known before-hand then a single measurement of duration $t_1$ yields $\Delta \hat{\omega} = \Delta \hat{\varphi}(t_1)/t_1$. Setting $t_1=t_{\pi} \equiv \pi/\Omega$, i.e. the time to perform a $\pi$-rotation on the atom, we have $\Delta \hat{\varphi} \geq 1$ and thus, $\Delta \hat{\omega} \geq 1/t_1$. By the 1TF conjecture and noting that the same uncertainty limit restricts Ramsey spectroscopy in atomic clocks [13], such a measurement is optimal, and no other single measurement provides lower frequency uncertainty in the same time. More precisely, if $t_1$ comprises a control $\pi/2$-rotation followed by a signal $\pi/2$-rotation, then the measurement provides an estimate of $\varphi(t_1/2)$, i.e. the signal phase at the mid-point of the experiment, thereby resulting in a factor of two worse uncertainty. However as information on two parameters is obtained, the signal phase and frequency, information theory constrains the uncertainty to be twice that obtained by single parameter estimation. Importantly, although optimal for two parameter estimation, such a measurement does not exceed the T-E limit, thus the requirement for multiple measurements.

By repeated estimation of $\varphi$ using signal and control qubit rotations (each of duration $t/2$) and population readout (Fig. 3a), a time-trace forming an $R$-point data-set $ \left\lbrace x[1], x[2], \ldots , x[R] \right\rbrace$ is obtained, where Pr(1) for the $n^{th}$ data-point is
\begin{equation}
\mathrm{Pr}\left(x[n] = 1; \delta \right) \approx \sin \left[\frac{\Omega t}{2} \right] \cos \left[ \frac{\varphi_m(n)}{2}\right]^2,
\end{equation}
assuming that the signal is resonant during interaction, where $\varphi_m(n) = \varphi+\frac{(n-1)\delta \tau}{2}+\frac{\delta t}{2}$ is the phase at the mid-point of the $n^{th}$ measurement and $\tau = t+t_{RO}$ is the time between the $n^{th}$ and the $(n+1)^{th}$ measurement, including the time taken to readout the spin state, $t_{RO}$ (SI Note 4). For known $\tau, \omega_c$ and using $\omega = \omega_c-\delta$, each measurement allows estimation of $\omega$, where the minimum statistical uncertainty of the $n^{th}$ measurement $\Delta \hat{\omega}_n = \sqrt{\mathrm{Var}[\hat{\omega},n]}$ is given by the Cramer-Rao lower bound (CRLB, see ref. [14] for an overview and SI Note 4 for a derivation):
\begin{equation}
\mathrm{Var}[\hat{\omega},n] > \frac{3 - \cos \left[ \varphi_m(n) \right] + 2 \cos \left[ \frac{\varphi_m(n)}{2} \right]^2 \cos \left[ \Omega t \right]}{n^2 t^2 \sin \left[ \frac{\varphi_m(n)}{2} \right]^2 \sin \left[ \frac{ \Omega t}{2} \right]}.
\end{equation}
Evaluated at $t_\pi$, a frequency uncertainty of: $\Delta \hat{\omega} \geq \frac{2 \Omega}{n \pi}$ is obtained. As a result, we conclude that the minimum uncertainty from the entire dataset is given when every measurement provides one bit of information on the signal, taking $t_\pi$, time. For times shorter than $t_\pi$, the frequency uncertainty improves according to: $\Delta \hat{\omega} \geq \frac{1}{4 \Omega t^2}$ (see SI Note 4). Similar quadratic scaling behavior has led researchers to suggest that a frequency uncertainty $\frac{1}{\Omega T}$ below the T-E limit can be obtained from a single long measurement [15-17]. Here we use the 1TF conjecture to bypass such analysis. Alternatively one could reason that extending the interaction time leverages quantum coherence and therefore reduces to estimation of the atomic energy transition which should be avoided as it is limited by $1/T$.

Using:
\begin{equation}
\mathrm{Var}[\hat{\omega},R] = 1/\underset{n=1}{\overset{R}{\sum }}I(\omega,n),
\end{equation}
where $I(\omega,n) = 1/\mathrm{Var}[\hat{\omega},n]$ is the Fisher Information of the $n^{th}$ measurement [18], and evaluating Eq.\,(3) at $t_\pi$ we have:
\begin{equation}
\mathrm{Var}[\hat{\omega},R] > \frac{24}{t_\pi{}^2(R + 3R^2 + 2R^3)}, \,\,\,\,\,\,\,\,\,\, \mathrm{Var}[\hat{\omega},T] > \frac{24 t_\pi}{T(T + t_\pi)(2T+t_\pi)},
\end{equation}
assuming perfect and instantaneous readout. For large $R$, an uncertainty approximately $\sqrt{R}$ below the Heisenberg T-E limit can be obtained from a dataset containing $R = T/t_\pi$ points. Furthermore, for datasets containing a significant fraction of a period of $\delta$, the systematic error in $\hat{\omega}$ (arising from $\Delta\hat{\varphi}_{sys}$) goes to zero. Thus, by referencing the estimator to the readout frequency rather than the atomic phase, we have decoupled $\Delta\hat{\omega}$ from $\omega_0$ and without reducing the accuracy of frequency estimation.

As $\Delta\hat{\omega}$ is largely independent of $\delta_s$, $\Omega$, $\omega_0$ and $\varphi$, implementation of the technique is experimentally practical. In Fig. 3 we show the results of collecting a data-set where $\omega_s=1509224897.8$\,Hz, $\omega_c=1509246000.0$\,Hz, $\tau = 0.7520416666666667$\,$\mu$s and a least-squares fit is used to estimate the signal frequency. For measurement times of 100\,ms and 64\,seconds the power spectrum obtained with a fast fourier transform and a sinc fit to the maximum of the dataset is shown (Fig. 3b). A plot of the one standard deviation fit error to the signal frequency as a function of measurement time is shown (Fig. 3c) and compared to the T-E limit and state-of-the-art atomic clocks [5, 19]. The statistical error of $\Delta\hat{\varphi}_{stat}= 1.4 \times 10^{-5}$\,Hz at 64\,seconds is more than a thousand below the Heisenberg T-E limit, approximately $10^4$ better than single ion atomic clocks (Fig. 3c green) and also significantly below lattice clocks containing 1,000 atoms [20]. Notably, the NV center coherence time used in this experiment is orders of magnitude shorter than the atomic clocks. As the applied signal frequency is experimentally known, from $\Delta\hat{\varphi}_{sys}=\sqrt{(\hat{\omega}^2-\omega^2)}$, we find that additional systematic errors in the fit (error-bars in Fig. 3) are negligible. Estimation of the signal frequency however requires knowledge of $\tau$, and thus a precise clock. To that end, a GPS-disciplined quartz oscillator was used for timing, with a systematic error less than 1 part in $10^{10}$, due to atmospheric shifts [21]. Accounting for systematic timing errors, the absolute frequency uncertainty remains below the T-E limit for several seconds (Fig. 3d).

Fig. 3 does not demonstrate construction of a better atomic clock, where the figure of merit is fractional frequency uncertainty. Rather, only the absolute frequency precision is superior. The differences to atomic clock time standards where a higher carrier frequency is desired, as well as the interplay of rotation speed on precision are highlighted in Fig. 3e where the same technique was applied to the $^{14}$N nuclear spin of the NV center with $\omega_s=5090821.4$\,Hz, $\omega_c=5090800.0$\,Hz, $\tau=14.33958333333333$\,$\mu$s. Due to the lower gyromagnetic ratio a minimum $t_\pi = 5.25$\,$\mu$s could be achieved, which along with additional readout overheads results in higher experimental $\Delta\hat{\varphi}_{stat}$ and theoretical CRLB than obtained with the electron spin. However the lower carrier frequency reduces systematic timing errors, resulting in an overall better frequency uncertainty. Similar frequency estimation techniques have recently been reported [22, 23] without such error analysis or explicit comparison to the T-E limit.

In the methods we prove that, given the 1TF and quantum speed limit [24, 25], no series of measurements with a single atom can outperform Eq. (5). The bound implies that either \emph{i)} a sensor composed of $N$ atoms can provide a better than linear uncertainty reduction, or \emph{ii)} access to a single atom and $NT$ time provides a lower uncertainty than access to $N$ atoms and $T$ time. This observation provides strong motivation to analyse the limits to frequency estimation with $N$ atoms.

From the 1TF conjecture, we have the uncertainty of a single measurement on an entangled ensemble is: $\Delta\hat{\omega}_{ent} \geq 1/T$, independent of $N$, since the ensemble is treated as single system. Thus any benefit from entanglement must arise from a faster rotation speed allowing more samples per unit time to be obtained. From the Margolus-Levitin [24] and Mandelstam-Tamm [26] inequalities we have $t_{\pi,N} \geq t_\pi/N$ where $t_{\pi,N}$ is the time to rotate the entangled system into an orthogonal state. Accounting for faster evolution, a variance at least $\frac{24 t_\pi}{N T(T + t_\pi/N)(2T+t_\pi/N)}$ could be obtained with an entangled system. However, a lower uncertainty is obtained by independently estimating the frequency with each atom and performing $R=T/t_\pi$ samples in $T$. Thus we propose the following limit for frequency estimation with access to $N$ atoms:
\begin{equation}
\mathrm{Var}[\hat{\omega},R,N] > \frac{24}{N t_\pi{}^2(R + 3R^2 + 2R^3)}, \,\,\,\,\,\,\,\,\,\, \mathrm{Var}[\hat{\omega},T,N] > \frac{24 t_\pi}{N T(T + t_\pi)(2T+t_\pi)}.
\end{equation}
I.e. an uncertainty improvement of $1/\sqrt{N}$ compared to the single atom case.

Explicit consideration of entangling schemes are useful as a check of this bound and to provide physical insight. The entangled NOON state: $\frac{1}{\sqrt{2}} \left(\ket{0}+\ket{1} \right)^{\bigotimes N}$ for example, saturates the speed limit for evolution to an orthogonal state. With two-qubits, where readout and preparation is performed in the (computational) $z$-basis, we show that entanglement is disadvantageous. Starting from $\ket{00}$, two rotations by the control field taking time $t > 3 t_\pi/2$ prepares the NOON state. Interaction with the signal for a time $t_{\pi,2}=t_\pi /2$, before disentanglement with the control for $t > 3 t_\pi/2$ results in an output state $\ket{00}$ with probability $\left(1/2-1/2 \cos\left[2 \varphi\right]\right)^2$ (see SI Note 5), but takes more than three times longer than a measurement on the unentangled state. Similar analysis when preparation and readout of the spin is performed in the Bell basis instead of the computational basis, appears to provide an uncertainty better than Eq. (6), but we assert that correct application of energy conservation arguments such as in the single atom case leads to a lower probability of success in such experiments, and indeed is seen experimentally [27].

These results provide a metrological problem that can not be attacked with entanglement, irrespective of decoherence, and which yield experimentally testable predictions. The results also greatly limit the utility of quantum algorithms when applied to spectrum analysis, a major expected application of quantum computers [28]. In the following we analyse the performance of the quantum fourier transform (QFT) for frequency estimation and show that it is constrained by the T-E limit and is thus outperformed by classical signal processing algorithms.

Close connection between estimation of the signal phase and the QFT can be seen by noting that for $\varphi =\pi/2$, a $\pi/2$-rotation mediated by the signal at time $t$, brings the spin to a state $\ket{\psi_{QFT}(t)} \approx \frac{1}{\sqrt{2}}\left(\ket{0}+ e^{i\left(\omega_0+\delta_s\right)t} \ket{1}\right)$ in the lab frame or $\ket{\psi_{QFT} (t)} \approx \frac{1}{\sqrt{2}}\left(\ket{0}+ e^{i\delta t} \ket{1}\right)$ in the frame rotating with the control frequency. With $N$ spins starting in $\ket{0}^{\otimes N}$, and allowing the $N^{th}$ spin to interact with the signal at time $t_N=\frac{T}{2^{N-1}}$  during $T$, $\delta$ is encoded into the QFT state of the register via $N$ single qubit gates (Fig. 4). Application of the inverse QFT through either entangling gates and register readout or the semi-classical algorithm and conditioned measurements [29, 30] allows estimation of $\delta$ with an uncertainty, $\Delta \hat{\delta} \approx 1/T$ [31] from a total of $2N$ gates, where readout operations are performed with the control field. For arbitrary $\varphi$, a two-fold higher uncertainty is obtained [32]. The protocol allows $N$ different frequencies to be discriminated with a Heisenberg limited precision, or an improved dynamic range [33] when compared to a single measurement on a single atom. However, if $R$ measurements during $T$ are performed with a single atom, then a frequency uncertainty $\sim \sqrt{R}$ better than the $N$-qubit QFT is obtained, albeit requiring more computational overhead for the fitting algorithm. Alternatively, a comparable frequency uncertainty is obtained by taking the time of the first $\ket{1}$ as the inverse of the estimator. As each unitary gate depends only on the signal power, we assert that this protocol performs a general QFT on an arbitrary multi-frequency signal if the maximum signal power is known. Interestingly, we have shown that an algorithm at the heart of many expected speed-ups in quantum computation can be performed without exploiting entanglement, since although the number of gates increases linearly with $N$, the timing of the encoding gates is exponentially spaced. The results also raise questions on the ability to simulate the dynamics of quantum systems by exchanging running time with physical resources. 

Finally, we introduce two principles which can be of use when characterizing metrological performance. The first is a response to an observation by N. Ramsey that information about the energy levels of an atom can be obtained over a duration where no interaction with the radiation field occurs [13]. Here, improved estimation of the frequency of the radiation field (but not the atomic energy levels) is obtained when the sensor and signal interact continuously as compared to separating their interaction with a free evolution time. Thus we propose that in general, parameter estimation is always improved by reducing the time that the signal and sensor do not interact. The second principle (of continuous optimal sensitivity), is that if a given state-vector does not provide optimal sensitivity at all times, then it can not be used to obtain optimal sensitivity. This principle disallows overheads where time spent preparing exotic states is recovered by ‘catching-up’ to or ‘overtaking’ less sensitive states. We believe both statements can be made rigorous, since the sensitivity on quantum states is a metric, known as statistical distance [34], and when evolution is parameterized  by the signal, then: \emph{i)} the statistical distance does not increase when no interaction occurs, and \emph{ii)} maximizing the statistical distance at some time, requires maximizing the statistical distance at all shorter times. If true these principles would rule-out entanglement as a route to achieving optimal sensitivity in general unless entanglement generation can be integrated into sensing without overhead. Whilst entanglement may be used to improve sensitivity in devices operating well below these limits, we suggest that those resources could be more effectively employed on classical techniques for noise reduction or signal enhancement. A direct application of these principles to the techniques presented here appears to result in contradictions at two timescales. For times short enough such that only a single measurement is performed, an uncertainty twice the optimal limit is obtained, whereas at long timescales atomic clocks can obtain a better uncertainty due to reduced systematic error. We resolve the first conflict by requiring any method that provides $\Delta \hat{\omega} \approx 1/T$ from a single measurement to take least $4 t_\pi$ time. The second conflict can be resolved noting that atomic clocks are optimized to estimate $\omega_0$ or time, and not the signal frequency, for which a clock is requisite.

Here we have demonstrated a single atom device operating with a precision orders of magnitude beyond previous theoretic limits, thus vastly improving upon the performance limits of quantum spectrometers. In doing so we have bounded the precision a quantum computer can estimate an unknown frequency dependent on running time. Our results have direct impact on the application of quantum computers and quantum algorithms to spectrum analysis, and on the benefit of entanglement for frequency estimation and quantum sensing in general.

\subsection{References}
1.	Bollinger, J.J., et al., Optimal frequency measurements with maximally correlated states. Physical Review A, 1996. 54(6): p. R4649-R4652.\\
2.	Giovannetti, V., S. Lloyd, and L. Maccone, Quantum-Enhanced Measurements: Beating the Standard Quantum Limit. Science, 2004. 306(5700): p. 1330-1336.\\
3.	Aharonov, Y., S. Massar, and S. Popescu, Measuring energy, estimating Hamiltonians, and the time-energy uncertainty relation. Physical Review A, 2002. 66(5): p. 052107.\\
4.	Leibfried, D., et al., Toward Heisenberg-Limited Spectroscopy with Multiparticle Entangled States. Science, 2004. 304(5676): p. 1476-1478.\\
5.	Sanner, C., et al., Optical clock comparison for Lorentz symmetry testing. Nature, 2019. 567(7747): p. 204-208.\\
6.	Aharonov, Y. and D. Bohm, Time in the Quantum Theory and the Uncertainty Relation for Time and Energy. Physical Review, 1961. 122(5): p. 1649-1658.\\
7.	Zwierz, M., C.A. Pérez-Delgado, and P. Kok, General Optimality of the Heisenberg Limit for Quantum Metrology. Physical Review Letters, 2010. 105(18): p. 180402.\\
8.	Gabor, D., Acoustical Quanta and the Theory of Hearing. Nature, 1947. 159(4044): p. 591-594.\\
9.	Cohen, L., Time-Frequency Analysis. 1995: Prentice Hall PTR, Englewood Cliffs, N.J.\\
10.	Balasubramanian, G., et al., Ultralong spin coherence time in isotopically engineered diamond. Nature Materials, 2009. 8(5): p. 383-387.\\
11.	Doherty, M.W., et al., Theory of the ground-state spin of the NV$^-$ center in diamond. Physical Review B, 2012. 85(20): p. 205203.\\
12.	Jacques, V., et al., Dynamic Polarization of Single Nuclear Spins by Optical Pumping of Nitrogen-Vacancy Color Centers in Diamond at Room Temperature. Physical Review Letters, 2009. 102(5): p. 4.\\
13.	Ramsey, N.F., A Molecular Beam Resonance Method with Separated Oscillating Fields. Physical Review, 1950. 78(6): p. 695.\\
14.	Kay, S.M., Fundamentals of statistical signal processing, volume I: estimation theory. 1993.\\
15.	Pang, S. and A.N. Jordan, Optimal adaptive control for quantum metrology with time-dependent Hamiltonians. Nature Communications, 2017. 8: p. 14695.\\
16.	Gefen, T., F. Jelezko, and A. Retzker, Control methods for improved Fisher information with quantum sensing. Physical Review A, 2017. 96(3): p. 032310.\\
17.	Naghiloo, M., A.N. Jordan, and K.W. Murch, Achieving Optimal Quantum Acceleration of Frequency Estimation Using Adaptive Coherent Control. Physical Review Letters, 2017. 119(18): p. 180801.\\
18.	Fisher, R.A., Theory of Statistical Estimation. Mathematical Proceedings of the Cambridge Philosophical Society, 2008. 22(5): p. 700-725.\\
19.	Diddams, S.A., et al., An Optical Clock Based on a Single Trapped 199Hg+ Ion. Science, 2001. 293(5531): p. 825-828.\\
20.	McGrew, W.F., et al., Atomic clock performance enabling geodesy below the centimetre level. Nature, 2018. 564(7734): p. 87-90.\\
21.	Trimble Thunderbolt. https://timing.trimble.com/wp-content/uploads/thunderbolt-e-gps-disciplined-clock-datasheet.pdf.\\
22.	Schmitt, S., et al., Submillihertz magnetic spectroscopy performed with a nanoscale quantum sensor. Science, 2017. 356(6340): p. 832-837.\\
23.	Meinel, J., et al., Heterodyne Sensing of Microwaves with a Quantum Sensor. arXiv preprint, 2020. arXiv:2008.10068 \\
24.	Margolus, N. and L.B. Levitin, The maximum speed of dynamical evolution. Physica D: Nonlinear Phenomena, 1998. 120(1): p. 188-195.\\
25.	Jones, P.J. and P. Kok, Geometric derivation of the quantum speed limit. Physical Review A, 2010. 82(2): p. 022107.\\
26.	Mandelstam, L. and I. Tamm, The Uncertainty Relation Between Energy and Time in Non-relativistic Quantum Mechanics, in Selected Papers, B.M. Bolotovskii, V.Y. Frenkel, and R. Peierls, Editors. 1991: Berlin, Heidelberg.\\
27.	Bernien, H., et al., Heralded entanglement between solid-state qubits separated by three metres. Nature, 2013. 497(7447): p. 86-90.\\
28.	Lloyd, S., M. Mohseni, and P. Rebentrost, Quantum principal component analysis. Nature Physics, 2014. 10: p. 631.\\
29.	Griffiths, R.B. and C.-S. Niu, Semiclassical Fourier Transform for Quantum Computation. Physical Review Letters, 1996. 76(17): p. 3228-3231.\\
30.	Chiaverini, J., et al., Implementation of the Semiclassical Quantum Fourier Transform in a Scalable System. Science, 2005. 308(5724): p. 997-1000.\\
31.	Nielsen, M.A. and I.L. Chuang, Quantum Computation and Quantum Information. 2000, Cambridge, U. K.: Cambridge University Press.\\
32.	Higgins, B.L., et al., Entanglement-free Heisenberg-limited phase estimation. Nature, 2007. 450(7168): p. 393-396.\\
33.	V. Vorobyov, et al., Quantum Fourier transform for quantum sensing arXiv preprint, 2020. arXiv:2008.09716 \\
34.	Wootters, W.K., Statistical distance and Hilbert space. Physical Review D, 1981. 23(2): p. 357-362.

\begin{figure}[p]
\includegraphics[width=0.95 \textwidth]{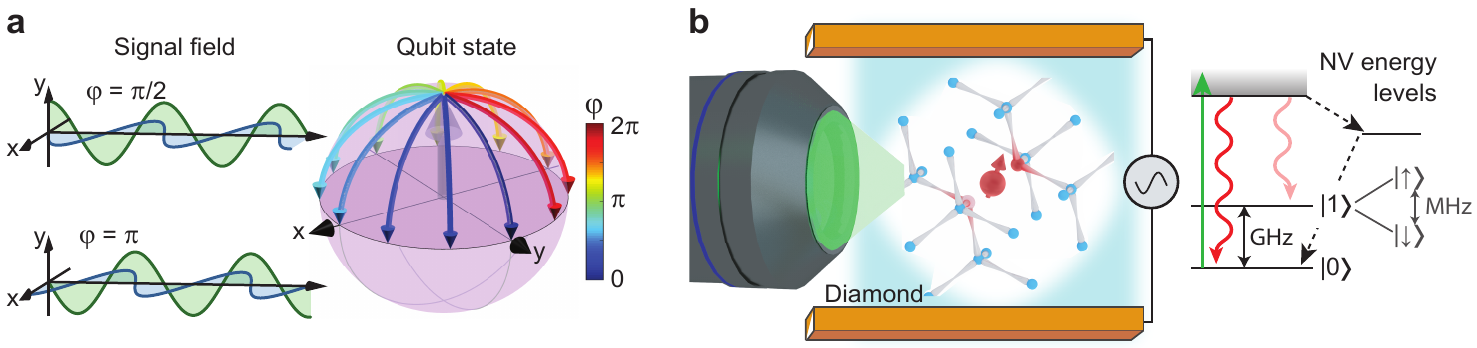}
\protect\caption{\textbf{Phase and frequency estimation with a single atom and experimental schematic.} (a) Interaction between an atom in state $\ket{0}$ and a near-resonant signal for a duration required to perform a $\pi/2$-rotation, maps the signal phase to the atomic state on the equator of the Bloch sphere. (b) The interaction in (a) is realised experimentally by applying an oscillating field to coplanar striplines along a diamond that contains a single nitrogen-vacancy (NV) center. The spin-state of the NV center is readout optically with a confocal microscope. Near-resonant rotation of the NV electron spin between $\ket{0} \leftrightarrow \ket{1}$ levels is performed at gigahertz frequencies. Megahertz frequency fields are used to drive the NV nuclear spin in the $\ket{1}$ manifold between $\ket{\uparrow} \leftrightarrow \ket{\downarrow}$.}
\label{fig1}
\end{figure}
 
\begin{figure}[p]
\includegraphics[width=0.95 \textwidth]{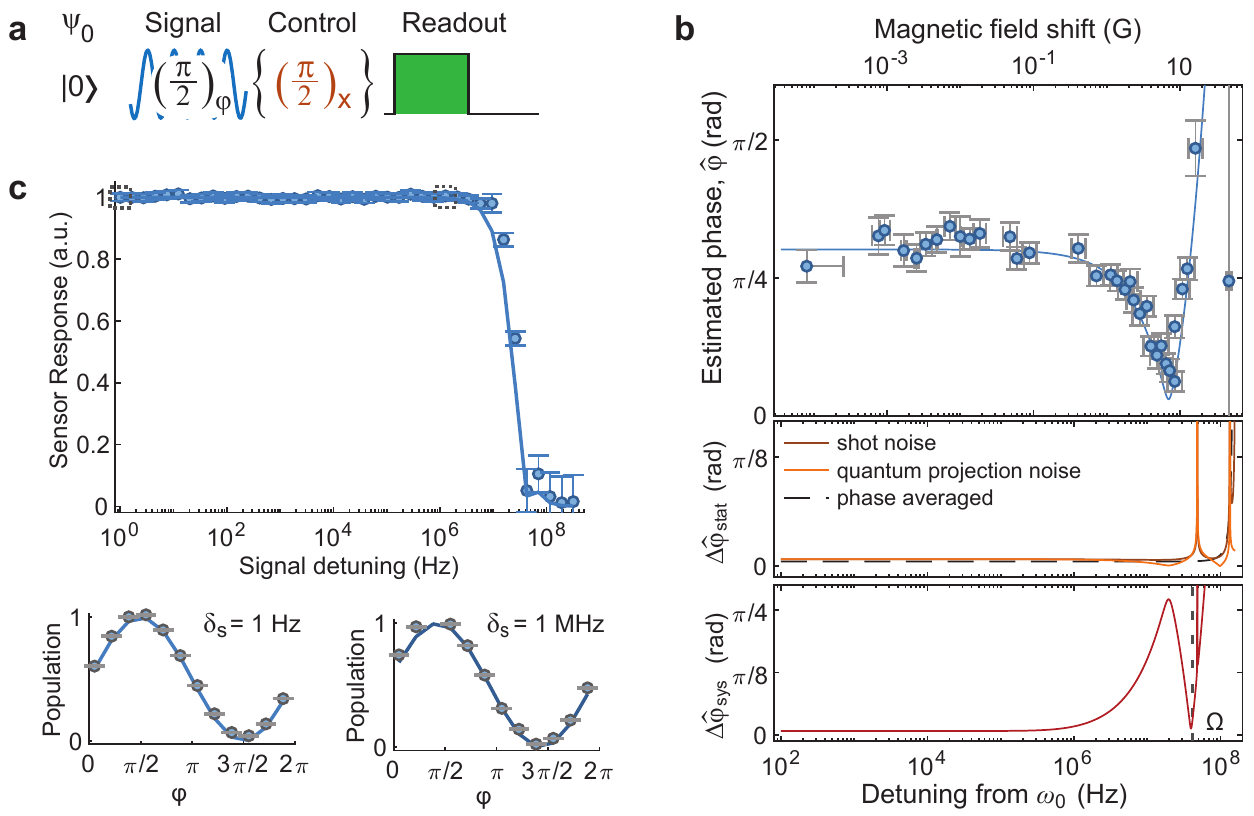}
\protect\caption{\textbf{Atomic energy independent and noise resilient quantum parameter estimation.} (a) The phase $\varphi$ of a near resonant signal is estimated with a $\pi/2$-rotation on the NV spin starting from state $\ket{0}$, followed by a $\pi/2$-rotation with a control field and population readout. (b) Phase estimation of signal with $\varphi = 0.95$, $\omega = 1.511406$\,GHz, using the protocol in (a) when the control field has the same frequency as the signal, and the external magnetic field is varied to detune the atomic frequency from resonance. (Upper panel) The estimated signal phase $\hat{\varphi}$ as a function of the external magnetic field and detuning from atomic resonance $\omega_0$, obtained from $10^4$ experiments reading out the NV spin state in $\ket{1}$ at each magnetic field value. (Middle panel) Theoretical statistical error in $\hat{\varphi}$ due to quantum projection noise (orange) and photon shot noise (brown) for $\varphi=0.95$, and the average error due to photon shot noise for $\varphi$ in the range $0\,-\,2\pi$ (black dotted) as a function of detuning from resonance (see SI Note 3 for fit equations). (Lower panel) Theoretical systematic error in $\hat{\varphi}$ (red) as a function of detuning from resonance, compared to the signal Rabi frequency $\Omega$ (black dotted). (c) The NV response to $\varphi$, measured by the population difference for $\varphi$ over the range $0\,-\,2\pi$, as a function of the signal frequency, using the protocol in (a). (Upper panel) The population difference obtained from $10^5$ experiments where NV transition frequency and control field are set constant to 1.5103\,GHz and the signal frequency is varied. Solid line is a sinc fit to the data. (Lower panel) Exemplary NV population readout curves as a function of $\varphi$ for signal detunings of 1\,Hz and 1\,MHz. Error bars are one standard deviation.}
\label{fig2}
\end{figure}

\begin{figure}[p]
\includegraphics[width=0.95 \textwidth]{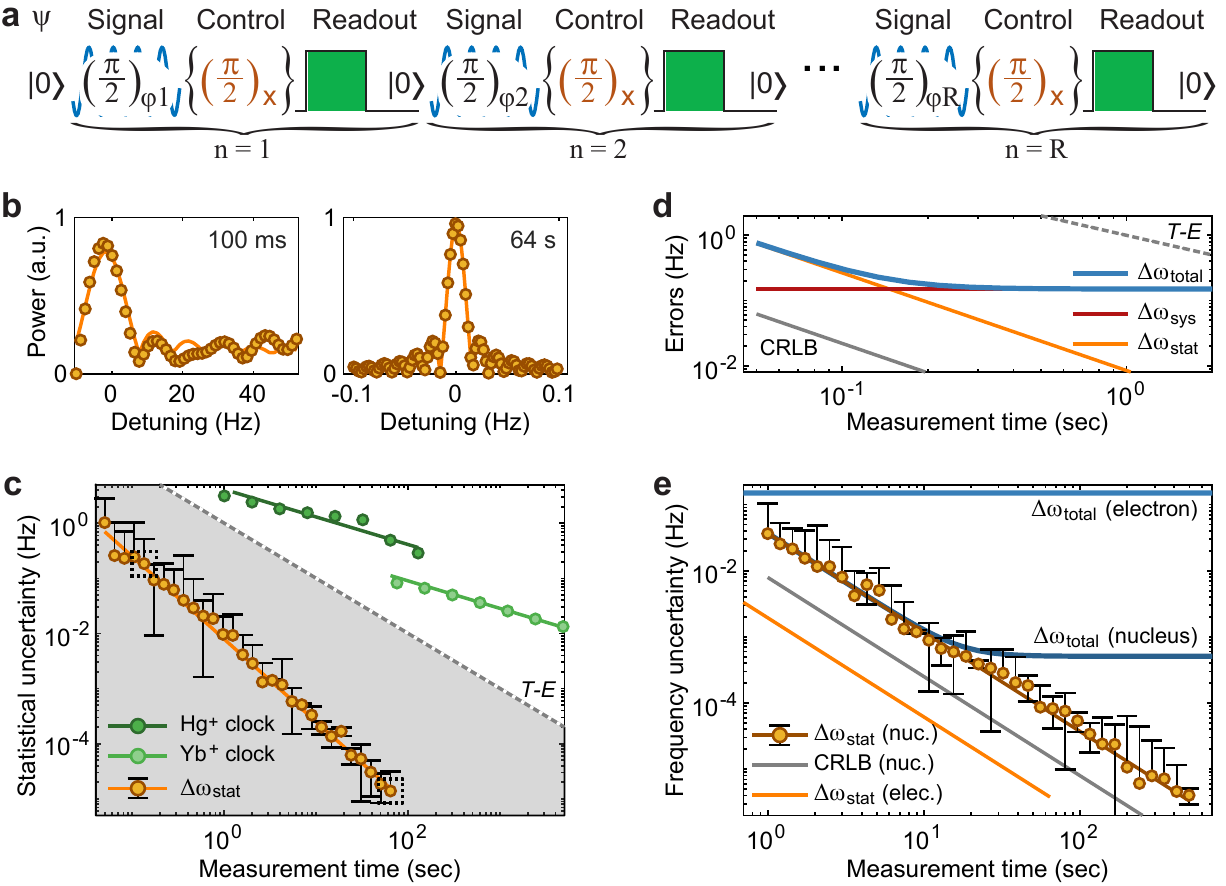}
\protect\caption{\textbf{Frequency estimation beyond the Heisenberg uncertainty limit with a single atomic sensor.} (a)The experimental protocol is a series of phase estimation procedures as shown in Fig. 2a from which the frequency is obtained as the phase change per unit time. (b) Fast fourier transform of the signal power spectrum for a field oscillating at 1509224897.8\,Hz recorded with a single NV center for 100 ms (left) and 64 seconds (right), and sinc fits. (c) Statistical uncertainty in the estimated field frequency (yellow) obtained from the least-squares fit error as a function of measurement time. Comparison to the Heisenberg T-E uncertainty limit (gray dotted), single Hg$^+$ ion (dark green) and Yb$^+$ ion (light green) atomic clocks (adapted from refs. [19] and [5] respectively). (d) Error budget showing statistical error (yellow), systematic error (red) and total error (blue) in estimation of the signal frequency as a function of time. Comparison to the Cramer-Rao lower bound from Eq. (5) (gray solid) and the Heisenberg T-E limit (gray dotted). (e) Frequency estimation with a single nuclear spin.  Statistical uncertainty in the estimated field frequency (brown) for a field oscillating at 5090821.4\,Hz, measured with the $^{14}$N nuclear spin of the NV center, and total error (dark blue). Comparison to the Cramer-Rao lower bound from Eq. (5) for the nuclear spin (gray solid), statistical error with the NV electronic spin (yellow) and total error with the NV electronic spin (blue). Error bars are the difference between the estimated and known frequency $|\omega-\hat{\omega}|$.}
\label{fig3}
\end{figure}

\begin{figure}[p]
\includegraphics[width=0.95 \textwidth]{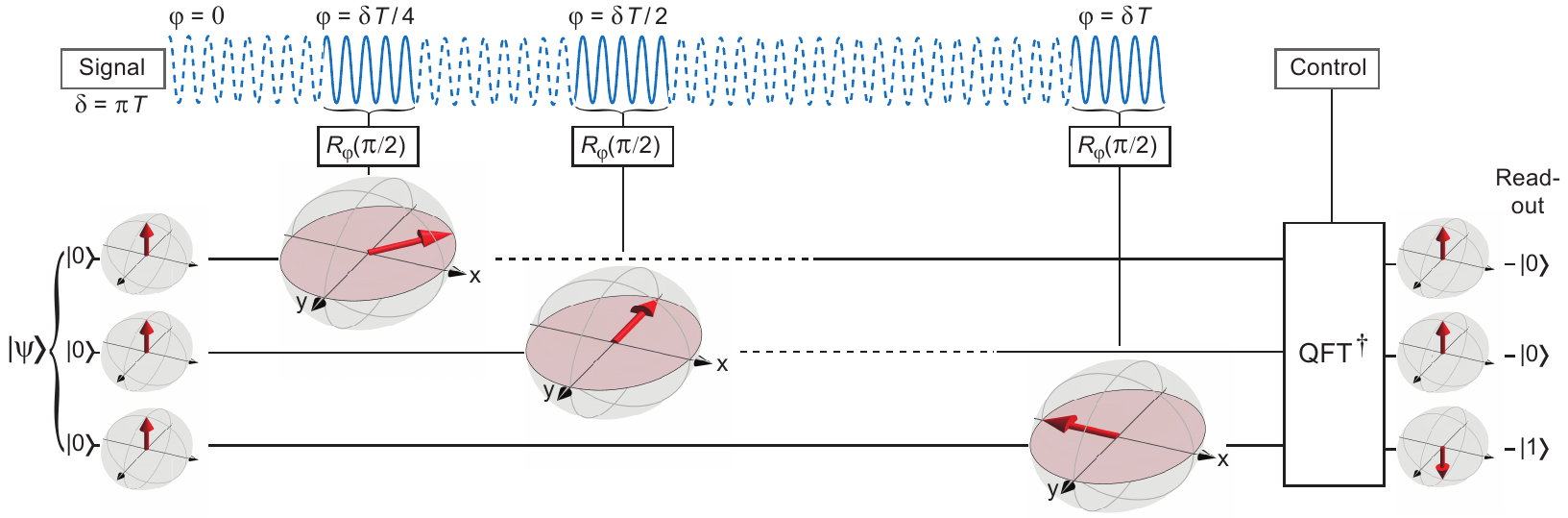}
\protect\caption{\textbf{The quantum fourier transform for frequency estimation.} An $N$-qubit quantum fourier transform (QFT) that encodes the signal frequency into the state of an $N$ qubit register is provided by using the signal to apply a $\pi/2$-rotation to the $N^{th}$ qubit at time $\frac{T}{2^{N-1}}$, where $T$ is the total measurement time. The protocol is shown explicitly for a 3 qubit register in a frame rotating with the control (and qubit) frequency and a signal detuned from the control field by $\delta = \pi T$. An inverse QFT performed at time $T$ brings the register to a state (here $\ket{001}$) from which the estimated signal frequency can be obtained as a binary expansion. Note a signal phase shift of $\pi/2$ with respect to the main text, and left-hand coordinate system are used for illustrational clarity and correspondence with other definitions of the QFT [31].}
\label{fig4}
\end{figure}

\newpage

\section*{Supplementary Information for frequency measurements beyond the Heisenberg time-energy limit with a single atom}

\subsection*{SI Note 1: The single measurement time-frequency uncertainty conjecture (1TF)}
Here we conjecture that no individual measurement on a single system can provide a frequency uncertainty less than the inverse of the measurement time, with respect to an unknown frequency $\omega$. For rigour, we should precisely define what we mean by `unknown', and  `frequency' in this context. We shall assume that the \emph{a priori} uncertainty in the frequency is at least $2/T$, where $T$ is the measurement time. When the unknown frequency refers to the Larmor frequency, $\omega_{\text{0}}$ of a single spin, the conjecture has close connection to the Heisenberg time-energy uncertainty principle.  However, here we specifically consider frequency estimation of classical fields of the form shown in Eq.\,(1). Thus we consider estimation of $\omega$ in a Hamiltonian equivalent to Eq.\,(1). In this respect the conjecture more closely mirrors the Gabor-Fourier limit.

The conjecture holds that when multiple atoms are entangled, such that the measurement outcomes on any atom can be perfectly correlated with the measurement outcomes on the remaining atoms, then they should be treated as a single system. In this respect the conjecture differs significantly from the Heisenberg limit for estimating the Larmor frequency of $N$ atoms: $\Delta \hat{\omega}_{\text{0}}\geq 1/(NT)$. With a difference of precisely a factor of $N$.

In the following we outline several arguments in favor of the 1TF conjecture providing the bound on frequency estimation of classical fields. We subsequently provide several counter-arguments. On the balance, we believe that current evidence indicates that the 1TF conjecture is true. Their discussion is organized below as follows: arguments for the 1TF conjecture; \emph{i)} with a single atom, \emph{ii)} with multiple atoms. Arguments against the 1TF conjecture; \emph{iii)} with a single atom, \emph{iv)} with multiple atoms.\\

\emph{The case for the 1TF: A single measurement of $\Delta \hat{\omega}\geq 1/T$ with a single atom}
\begin{itemize}
\item The Gabor-Fourier limit appears to preclude a better frequency uncertainty from a single sample.
\item If such a measurement could be used to estimate the Larmor frequency of the sensor, then achieving an uncertainty beyond this limit would violate the Heisenberg time-energy principle.
\item For a modified form of Eq.\,1, so that the classical field to be estimated comprises just a single photon, the 1TF is congruent/equivalent to the Heisenberg time-energy uncertainty principle.
\item All current experimental evidence (in particular, from atomic clocks) is in accord with this limit.
\end{itemize}

\emph{The case for the 1TF: A single measurement of $\Delta \hat{\omega}\geq 1/T$ with $N$ entangled atoms}
\begin{itemize}
\item The Gabor-Fourier limit appears to preclude better frequency estimation from a single sample.
\item The advantage of entanglement can often be emulated by replacing all atoms with a single system of higher magnetic moment. This case reduces to estimation of an $N$-fold higher frequency with a single atom, but the uncertainty in estimating the frequency of this field should also obey the 1TF.
\item Theoretical proposals that use an entangled $N$ spin system only take advantage of a two-dimensional subspace. As only a single bit of information is obtained upon readout, the uncertainty reduction from the initial prior can at most, be a half.
\item The main text demonstrates a frequency estimation strategy wiht an uncertainty that improves better than linearly in time. As a result, $\Delta \hat{\omega} \geq 1/(NT)$ cannot be the ultimate limit, as this is surpassed at some finite time by performing multiple measurements on a single atom.
\item The main text also breaks the symmetry allowing linear time to be exchanged with a linear number of particles. For time and particle number to maintain equivalent simulation power would require the frequency uncertainty to improve super-linearly with number of atoms.
\item Most persuasively. In terms of experimental evidence, we are unaware of any experiment using entanglement that has demonstrated a frequency uncertainty beyond the 1TF conjecture.
\end{itemize}

\emph{The case against the 1TF: A single measurement of $\Delta \hat{\omega} < 1/T$ with a single atom}
\begin{itemize}
\item The conjecture as formulated appears imprecise, vague and clunky.
\item The 1TF conjecture places limits on the phase estimation of fields of unknown frequency. In particular the 1TF implies that the uncertainty in phase estimation cannot indefinitely follow Heisenberg limit (linear improvement).
\item Theoretical work deriving the Heisenberg limit for frequency estimation produces a limit of: $\Delta \hat{\omega}\geq T^{-2}$.
\item Approximate and analytic evolution with simultaneous fields appears to contradict the 1TF. I.e. when the control and signal fields are applied simultaneously to the qubit for a time $t=\pi/(2 \Omega)$, and the frequency and amplitude of the signal field is known, then a better phase estimation (by a factor of $2\pi$) is obtained compared to when the signal and control fields are applied consecutively. The dynamics cannot be solved analytically when the signal is not co-resonant with the control field, but the Magnus expansion and other approximations appear to yield an uncertainty better than the 1TF conjecture for interaction times longer than $t_{\pi}$. However, an hour in the laboratory can sometimes save a week with Mathematica, and experiments do not support this analysis.
\item With prior knowledge on the signal frequency, an uncertainty better than the 1TF can be obtained by discriminating between two known frequencies in less time than the inverse of their frequency difference (SI Ref 1).
\end{itemize}

\emph{The case against the 1TF: A single measurement of $\Delta \hat{\omega} < 1/T$ with $N$ entangled atoms}
\begin{itemize}
\item The 1TF contradicts decades of research into the limits of quantum metrology.
\end{itemize}

\subsection*{SI Note 2: System Hamiltonian and time-evolution of the state vector}

Here we describe the dynamics of a spin-1/2 particle that interacts, at separate times with one of two near-resonant fields. We want to estimate
the (unknown) frequency \(\omega _s\) of the signal field, with access to an additional control field of known frequency \(\omega _c\). For a spin
in a magnetic field \(B_z\), with Larmor frequency \(\omega _0\) = $\gamma $ \(B_z\), where $\gamma $ is the gyromagnetic ratio of the spin, the
semi-classical Hamiltonian describing the interaction from \(t_0\) to \(t\) with a field that is circularly polarized in the x-y plane is:

\begin{equation}
H\left(t_0,t\right)=\frac{\hbar \omega _0}{2}\sigma _z+\frac{\hbar \Omega }{2}\cos \left[\text{$\omega $t}+\varphi \left(t_0\right)\right]\sigma
_x+\frac{\hbar \Omega }{2}\sin \left[\text{$\omega $t}+\varphi \left(t_0\right)\right]\sigma _y,
\end{equation}
where \(\omega\) is the field frequency, \(\Omega\) the Rabi frequency of the atom-field interaction, \(\varphi \left(t_0\right)\) the initial phase
of the field, $\sigma_{x,y,z}$ the Pauli spin-matrices (defined in the lab frame), and $\hbar$ the reduced Planck constant.\\
In the following analysis, \(s,c\) subscripts refer to the signal, control fields respectively, and we define { }\(\omega _{s,c} =\omega _0+\delta
_{s,c}\) as their detuning from atomic resonance, with a relative detuning \(\delta\) between signal and control, \(\delta _s = \delta _c+\delta\).
We assume that signal and control have the same Rabi frequency \(\Omega\), which is taken to be real and positive, and that the \(x, y\) axes are
defined such that the control has an initial phase of 0. Therefore no subscripts are used for \(\Omega ,\varphi\) without ambiguity. As only the signal
phase depends on \(t_0\), for concision we drop \(t_0\) notation with the understanding that \(\varphi\) depends on the starting time of the experiment.
If \(\delta_{s,c}  < \Omega\), then we say the field is near-resonant. \\
We write the Hamiltonian in the form of Eq.\,(1), where the oscillating field has circular, rather than linear polarization, as it removes the need
to invoke the rotating wave approximation and allows an exact solution. This Hamiltonian also realises the most efficient transfer of the field energy
to the spin so that for a given Rabi frequency, the spin is driven such that it satisfies the quantum speed limit. However a circular field is not
necessary for the main results of the paper, and experiments were performed with linearly polarized fields. Note also that in this definition a signal
phase of $\pi $/2 is required to perform the correct mapping to a QFT state as defined in Nielsen and Chuang (31).

To solve for the unitary evolution of the spin under Eq.\,(1), we move into a frame rotating with the field frequency, so that the Hamiltonian becomes
time-independent. The transformation matrix is

\begin{equation}
\mathcal{T}_{\omega }(t)=e^{i \omega  t \left.\sigma _z\right/2}.
\end{equation}

The Hamiltonians in the frames rotating with frequency \(\omega _{s,c}\) are then

\begin{equation}
\begin{array}{cc}
H_{\omega _s}^R=-\hbar  \delta _s\left.\sigma _z\right/2+\hbar  \Omega \left.\left(\cos [\varphi ]\sigma _x+\sin [\varphi ]\sigma _y\right)\right/2 \\
H_{\omega _c}^R=-\hbar  \delta _c\left.\sigma _z\right/2+\hbar  \Omega  \left.\sigma _x\right/2, \\
\end{array}
\end{equation}
where the \(\sigma _{x,y}\) spin-matrices now depend on the rotating frame. Note in the rotating frame, estimation of the field frequency is equivalent
to estimating the qubit energy. Unitary evolution of the state vector in the signal rotating frame is then

\begin{equation}
U^R\left(t,\delta _s,\varphi \right) = \frac{1}{\Omega _s}\left(
\begin{array}{cc}
 \Omega _s\text{cos}\left[t \left.\Omega _s\right/2\right]+i \delta _s \text{sin}\left[t \left.\Omega _R\right/2\right] & -i \Omega  e^{-i \varphi
} \text{sin}\left[t \left.\Omega _s\right/2\right] \\
 -i \Omega  e^{i \varphi } \text{sin}\left[t \left.\Omega _s\right/2\right] & \Omega _s\text{cos}\left[t \left.\Omega _s\right/2\right]-i \delta
_s \text{sin}\left[t \Omega _s\right] \\
\end{array}
\right),
\end{equation}
where \(\Omega _s=\sqrt{\delta _s{}^2+\Omega ^2}\), and evolution under control can be obtained by substituting \(\delta _s\to \delta _c\), \(\varphi
\to 0\). Importantly, when switching between rotating frames, the state vector must be transformed to maintain consistent evolution in all reference
frames. The state vector in the lab frame is obtained by applying the transformation \(\mathcal{T}_{\omega }(t)^{-1}\) which corresponds to a rotation
around the z-axis. When switching from control to signal frame, we rotate through an angle \(t_c \delta\) where \(t_c\) is the time spent in the
control frame, alternatively we can simply update the signal phase by an amount \(t_c \delta\) without transforming the state vector. Furthermore,
if we only readout the spin in a transformation-independent basis (e.g. $\{$\(|0\rangle\),\(|1\rangle\)$\}$ as is performed here), then fully consistent
results are obtained simply by updating the signal or control phase and there is no need to transform any operators. Approximating a $\pi $/2-pulse
from the signal as a perfect $\pi $/2 rotation, we can write the qubit state in the different rotating frames after a $\pi $/2 rotation as:

\begin{equation}
\begin{array}{cc}
|\psi (t)\rangle _S=\frac{1}{\sqrt{2}}\left(|0\rangle -i e^{i \varphi }|1\rangle \right)=\frac{1}{\sqrt{2}}\left(|0\rangle +e^{i (\varphi -\pi /2)}|1\rangle
\right)   &  (\text{signal frame}) \\
|\psi (t)\rangle _L=\frac{1}{\sqrt{2}}\left(|0\rangle -i e^{i \varphi }|1\rangle \right)=\frac{1}{\sqrt{2}}\left(|0\rangle +e^{i \left(\varphi -\pi
/2+\left(\omega _0+\delta _s\right)t\right)}|1\rangle \right)  &  (\text{lab frame}) \\
|\psi (t)\rangle _C=\frac{1}{\sqrt{2}}\left(|0\rangle -i e^{i \varphi }|1\rangle \right)=\frac{1}{\sqrt{2}}\left(|0\rangle +e^{i (\varphi -\pi /2+\delta
 t)}|1\rangle \right)  & (\text{control frame}) \\
\end{array}
\end{equation}

For one sequential application of the control and signal field, each for time \(t/2\), the unitary is given by
\begin{equation}
U_{\text{seq}}^R\left(t,\delta _s,\delta _c,\varphi \right) = U^R\left(t/2,\delta _s,\varphi +\delta  t/2\right).U^R\left(t/2,\delta _c\right)
\end{equation}

For a spin initially in the state \(|0\rangle\), the probability to measure the spin in state \(|1\rangle\) after evolution under \(U_{\text{seq}}^R\left(t,\delta_s,\delta _c,\varphi \right)\) is
\begin{equation}
\begin{split}
\Pr & \left(1; t,\delta _s,\delta _c,\varphi \right) =\left(\Omega^2 /\left(2 \Omega_c^2 \Omega_s^2\right)\right) \\
& (\text{sin}\left[\left(t\Omega _s\right)/4\right]^2 (\delta^2+\Omega ^2+\Omega^2 \text{cos}\left[\left(t \Omega_c\right)/2\right]+2 \delta  \delta_c+3 \delta_c^2 -\delta_s \\
& \left(4 \text{cos}\left[\varphi_m\right] \text{sin}\left[\left(t \Omega_c\right)/4\right]^2 \delta_c+\text{cos}\left[\left(t\Omega_c\right)/2\right] \delta_s+2 \text{sin}\left[\varphi_m\right] \text{sin}\left[\left(t \Omega_c\right)/2\right] \Omega_c\right)) \\
& +\text{sin}\left[\left(t\Omega_s\right)/2\right] \left(-2 \text{sin}\left[\varphi_m\right] \text{sin}\left[\left(t\Omega_c\right)/4\right]^2 \delta_c+\text{cos}\left[\varphi_m\right] \text{sin}\left[\left(t \Omega_c\right)/2\right] \Omega_c\right) \Omega _s + \\
& 2 \text{cos}\left[\left(t \Omega _s\right)/4\right]^2 \text{sin}\left[\left(t \Omega_c\right)/4\right]^2\Omega_s^2),
\end{split}
\end{equation}
where \(\varphi _m=\varphi +\frac{\delta  t}{2}\) is the signal phase at the middle of the sequence (after the application of the control field),
and by symmetry the probability remains the same when the order of the control and signal fields are reversed. If the control field is resonant with
the spin ($\delta_s $ = 0), then we have
\begin{equation}
\begin{split}
\Pr \left(1;t,\delta _s,0,\varphi _m\right) & =\frac{1}{2 \Omega _s^2} (\left(-\delta \Omega \text{cos}\left[\varphi_m-\frac{t \Omega}{2} \right]+\left(-\delta^2+\Omega^2\right) \text{cos}\left[\frac{t \Omega }{2}\right]+\delta  \Omega  \text{cos}\left[\varphi_m+\frac{t \Omega}{2}\right]\right) \\
& \text{sin}\left[\frac{t \Omega_s}{4}\right]^2+\Omega  \text{cos}\left[\varphi _m\right] \text{sin}\left[\frac{t \Omega }{2}\right] \text{sin}\left[\frac{t\Omega _s}{2}\right] \Omega_s+\\
& \left(2 \text{cos}\left[\frac{t \Omega _s}{4}\right]^2 \text{sin}\left[\frac{t \Omega }{4}\right]^2+\text{sin}\left[\frac{t\Omega _s}{4}\right]^2\right) \Omega _s^2).
\end{split}
\end{equation}
When \(U_{\text{seq}}^R\) is applied for a time \(t = \pi /\Omega\), i.e. long enough to perform a $\pi $-rotation on the spin, then
\begin{equation}
\begin{split}
\Pr & \left(1; \pi /\Omega ,\delta _s,\delta _c,\varphi _m\right)=\frac{\Omega ^2}{2 \Omega _c^3 \Omega _s^2} \\
& (-\text{sin}\left[\frac{\Pi _s}{2}\right]^2 \Omega _c (-\delta ^2-\Omega ^2-2 \delta  \delta _c-3 \delta _c^2+\text{cos}\left[\Pi _c\right] \left(-\Omega +\delta _s\right) \left(\Omega+\delta _s\right)+\\
& 2 \delta _s \left(2 \text{cos}\left[\varphi _m\right] \text{sin}\left[\frac{\Pi_c}{2}\right]^2 \delta _c+\text{sin}\left[\Pi_c\right] \text{sin}\left[\varphi _m\right] \Omega _c\right))+\\
& \text{sin}\left[\frac{\Pi _c}{2}\right] \text{sin}\left[\Pi _s\right] \left(\text{cos}\left[\varphi_m-\frac{\pi  \Omega _c}{4 \Omega }\right] \Omega _c \left(-\delta _c+\Omega _c\right)+\text{cos}\left[\varphi _m+\frac{\pi  \Omega _c}{4 \Omega}\right] \left(\Omega^2+\delta _c \left(\delta _c+\Omega _c\right)\right)\right) \Omega_s+\\
& 2 \text{cos}\left[\frac{\Pi_s}{2}\right]^2 \text{sin}\left[\frac{\Pi_c}{2}\right]^2 \Omega _c \Omega _s^2),
\end{split}
\end{equation}
where \(\Pi _s=\frac{\pi  \Omega _s}{2 \Omega },\Pi _c=\frac{\pi  \Omega _c}{2 \Omega }\). These expressions can be simplified in the special cases
that the signal and control fields have the same frequency with a detuning \(\delta _s\) from atomic resonance:
\begin{multline}
\Pr \left(1;t,\delta _s,\delta _s,\varphi \right) = \\
\frac{\Omega ^2 \text{sin}\left[\frac{t \Omega _s}{4}\right]{}^2 \left(4 \Omega ^2 \text{cos}\left[\frac{\varphi}{2}\right]^2 \text{cos}\left[\frac{t \Omega _s}{4}\right]{}^2+2 \left(1+\text{cos}[\varphi ] \text{cos}\left[\frac{t \Omega _s}{2}\right]\right)\delta_s^2-2 \text{sin}[\varphi ] \text{sin}\left[\frac{t \Omega _s}{2}\right] \delta _s \Omega _s\right)}{\Omega _s^4}
\end{multline}
and when signal and control are perfectly resonant to the spin:
\begin{equation}
\Pr (1;t,0,0,\varphi ) = \text{cos}[\varphi /2]^2 \text{sin}[t \Omega /2]^2.
\end{equation}
For two $\pi $/2-pulses, Eq.\,(11) recovers the result of atomic clocks using Ramsey spectroscopy:
\begin{equation}
\Pr (1;\pi /\Omega ,0,0,\varphi ) = \text{cos}[\varphi /2]^2.
\end{equation}

\subsection*{SI Note 3: Error in estimating the initial signal phase $\varphi $}

The probability to readout the NV center in \(|1\rangle\) after application of a $\pi $/2 pulses, in turn by two fields of the same frequency
(detuned from resonance by \(\delta _s\)) and a phase relationship of $\varphi $ is:
\begin{multline}
\Pr \left(1;\pi /\Omega ,\delta _s,\delta _s,\varphi \right) = \\
\frac{\Omega ^2 \text{sin}\left[\frac{\Pi _s}{2}\right]{}^2 \left(2 \delta _s{}^2+\Omega^2+\Omega ^2 \text{cos}[\varphi ]+\text{cos}\left[\Pi _s\right] \left(\Omega ^2+\left(2 \delta _s{}^2+\Omega ^2\right) \text{cos}[\varphi ]\right)-2
\delta _s \Omega _s \text{sin}\left[\Pi _s\right] \text{sin}[\varphi ]\right)}{\Omega _s{}^4}.
\end{multline}
From Eq.\,(13), denoting \(\Pr (1;\text{...})\equiv P\), the statistical error in estimating $\varphi $ is at least:
\begin{equation}
\Delta \hat{\varphi }_{\text{stat}}\geq  1\left/\sqrt{\left(\frac{\partial P}{\partial \varphi }\right)^2\frac{1}{P (1-P)}}\right.,
\end{equation}
when only considering quantum mechanical projection noise. Inserting Eq.\,(13 ) into Eq.\,(14) and taking a series expansion of \(\Delta \hat{\varphi
}_{\text{stat}}\) around \(\delta _s=0\) gives
\begin{equation}
\Delta \hat{\varphi }_{\text{stat}}\approx  1+\frac{(-4+\pi )^2 \text{csc}\left[\frac{\varphi }{2}\right]^2 \delta _s{}^4}{32 \Omega ^4}+O\left[\delta
_s\right]{}^5.
\end{equation}
At $\varphi $ = $\pi $/2, averaged between the minimum and maximum, an error of \(1+\frac{(-4+\pi )^2\delta _s{}^4}{16 \Omega ^4}\) is obtained.
If the experimental noise is photon shot noise of magnitude \(\sigma _{\text{SN}}\), the statistical error is given by
\begin{equation}
\Delta \hat{\varphi }_{\text{stat}}\geq  1\left/\sqrt{\left(\frac{\partial P}{\partial \varphi }\right)^2\frac{1}{\sigma _{\text{SN}}{}^2}}.\right.
\end{equation}
When averaged over the range $\varphi $ $\in $ (0,2$\pi $), the error is
\begin{equation}
\left\langle \Delta \hat{\varphi }_{\text{stat}}\right\rangle =\frac{\sqrt{2\sigma _{\text{SN}}}\Omega _s^4}{\sqrt{\Omega ^4 \left(2 \Delta ^2+\Omega^2+\Omega ^2 \text{cos}\left[\Pi _s\right]\right)^2 \text{sin}\left[\frac{\Pi _s}{2}\right]^4}}
\end{equation}
which has a series expansion around \(\delta _s=0\) of
\begin{equation}
\left\langle \Delta \hat{\varphi }_{\text{stat}}\right\rangle \approx  2 \sqrt{2\sigma _{\text{SN}}}+\frac{(-4+\pi )^2 \delta _s{}^4\sqrt{\sigma
_{\text{SN}}}}{4 \sqrt{2} \Omega ^4}+O\left[\delta _s\right]{}^5
\end{equation}

with the same magnitude as the error due to projection noise when \(\sigma _{\text{SN}} = 1/8\).

To calculate the systematic error, we take the estimate of $\varphi $ from the state readout:
\begin{equation}
\hat{\varphi }=\text{cos}^{-1}[2(\Pr (1)-1/2)]
\end{equation}
with a systematic error given by: \(\Delta \hat{\varphi }_{\text{sys}}=\sqrt{\hat{\varphi }^2-\varphi ^2}\). A series expansion of \(\Delta \hat{\varphi}_{\text{sys}}\) around \(\delta _s=0\) gives
\begin{equation}
\Delta \hat{\varphi }_{\text{sys}} \approx \sqrt{\text{cos}^{-1}[\text{cos}[\varphi ]]^2-\varphi ^2}+\frac{2 \text{cos}^{-1}[\text{cos}[\varphi ]] \text{sin}[\varphi] \delta _s}{\Omega  |\text{sin}[\varphi ]| \sqrt{-\varphi ^2+\text{cos}^{-1}[\text{cos}[\varphi ]]^2}} + O\left[\delta _s\right]{}^2,
\end{equation}
which reduces to
\begin{equation}
\Delta \hat{\varphi }_{\text{sys}} \approx  \frac{2 \varphi  \delta _s}{\Omega }+O\left[\delta _s\right]{}^2
\end{equation}
if $\varphi $ is known to be in the range (0, $\pi $/2). We can perform the same error analysis for when the control is resonant to the qubit frequency, and the signal has an unknown and arbitrary detuning
\(\delta _s< \Omega\). In this case the probability to readout the NV center in \(|1\rangle\) after two $\pi $/2-pulses, one with a detuning \(\delta
_s\), and one with no detuning is

\begin{multline}
\Pr \left(1;\pi /\Omega ,\delta _s,0,\varphi \right)= \frac{2 \Omega _s{}^3-2 \delta _s \Omega  \Omega _s \text{sin}\left[\varphi +\frac{\pi \delta }{2 \Omega }\right]}{4 \Omega _s{}^3} + \\
\frac{-\Omega  \left(\Omega _s{}^2-\delta _s\Omega _s\right) \text{sin}\left[\frac{\pi  \delta +2 \varphi  \Omega -\pi  \Omega_s}{2 \Omega }\right]+\Omega  \left(\Omega ^2+\delta _s \left(\delta _s+\Omega _s\right)\right) \text{sin}\left[\varphi +\frac{\pi  \left(\delta +\Omega _s\right)}{2 \Omega }\right]}{4 \Omega _s{}^3}.
\end{multline}

Expression (22) can be made simpler by noting that in the following analysis the signal phase is arbitrary, allowing the substitution \(\varphi _m=\varphi
+\frac{\delta  \pi }{2 \Omega }\), where \(\varphi _m\) is the signal phase after the first $\pi $/2-pulse. Then we have 
\begin{equation}
\Pr \left(1;\pi /\Omega ,\delta _s,0,\varphi _m\right)=\frac{1}{2} \left(1-\frac{2 \delta _s \Omega  \text{sin}\left[\varphi _m\right] \text{sin}\left[\frac{\pi
 \Omega _s}{4 \Omega }\right]{}^2}{\Omega _s{}^2}+\frac{\Omega  \text{cos}\left[\varphi _m\right] \text{sin}\left[\frac{\pi  \Omega _s}{2 \Omega
}\right]}{\Omega _s}\right).
\end{equation}
Using Eq.\,(13), and taking a series expansion of \(\Delta \hat{\varphi }_{\text{stat}}\) around \(\delta _s=0\) gives
\begin{equation}
\Delta \hat{\varphi }_{\text{stat}}\approx  1+\frac{(-4+\pi )^2 \text{csc}\left[\varphi _m\right]{}^2 \delta ^4}{32 \Omega ^4}+O\left[\delta _s\right]{}^5.
\end{equation}
From Eq.\,(19) we have
\begin{equation}
\Delta \hat{\varphi }_{\text{sys}} \approx \sqrt{-\varphi _m{}^2+\text{cos}^{-1}\left[\text{cos}\left[\varphi _m\right]\right]{}^2}+\frac{\text{cos}^{-1}\left[\text{cos}\left[\varphi
_m\right]\right] \text{sin}\left[\varphi _m\right] \delta _s}{\Omega  |\text{sin}\left[\varphi _m\right]| \sqrt{-\varphi _m{}^2+\text{cos}^{-1}\left[\text{cos}\left[\varphi
_m\right]\right]{}^2}} + O\left[\delta
_s\right]{}^2.
\end{equation}
Which reduces to
\begin{equation}
\Delta \hat{\varphi }_{\text{sys}} \approx  \frac{2 \varphi _m \delta _s}{\Omega }-\frac{2 \varphi _m{}^2 \delta _s{}^2}{\Omega ^2}+O\left[\delta_s\right]{}^3.
\end{equation}
Thus we can see that the systematic and statistical errors are approximately the same when both signal and control fields have the same detuning
from atomic resonance, or when the control field is resonant to the atomic transition.

In addition to performing a series expansion of the statistical error, which shows an increase in error as a function of signal detuning, if we show
\begin{multline}
\left(\frac{\partial P\left(\delta _s=0\right)}{\partial \varphi }\right)^2\frac{1}{P \left(\delta _s=0\right)\left(1-P\left(\delta _s=0\right)\right)} > \\
\left(\frac{\partial P\left(\delta _s>0\right)}{\partial \varphi }\right)^2\frac{1}{P \left(\delta _s>0\right)\left(1-P\left(\delta _s>0\right)\right)}
\end{multline}
then we have that the estimator variance is minimised (and thus the atomic sensitivity to the signal phase is maximal) when the signal has zero detuning from atomic resonance. Solving Eq.\,(27) reduces
to
\begin{multline}
1>\\
\frac{\Omega ^2 \left(4 \delta _s \Omega _s \text{cos}\left[\frac{\pi  \Omega _s}{4 \Omega }\right] \text{sin}[2 \varphi ] \text{sin}\left[\frac{\pi
 \Omega _s}{4 \Omega }\right]^3+4 \delta _s{}^2 \text{cos}[\varphi ]^2 \text{sin}\left[\frac{\pi  \Omega _s}{4 \Omega }\right]^4+\Omega _s{}^2
\text{sin}[\varphi ]^2 \text{sin}\left[\frac{\pi  \Omega _s}{2 \Omega }\right]^2\right)}{\Omega _s{}^4+\Omega ^2 \left(4 \delta _s \Omega _s \text{cos}\left[\frac{\pi
 \Omega _s}{4 \Omega }\right] \text{sin}[2 \varphi ] \text{sin}\left[\frac{\pi  \Omega _s}{4 \Omega }\right]^3-4 \delta _s{}^2 \text{sin}[\varphi
]^2 \text{sin}\left[\frac{\pi  \Omega _s}{4 \Omega }\right]^4-\Omega _s{}^2 \text{cos}[\varphi ]^2 \text{sin}\left[\frac{\pi  \Omega _s}{2 \Omega
}\right]^2\right)}.
\end{multline}

Whilst unable to demonstrate analytically, from numerics we find that Eq.\,(27) is in general true.

\subsection*{SI Note 4: Fisher Information for estimating $\varphi $ and $\omega $}

The quantum Fisher Information (QFI) for a parameter $\varphi $, describes the amount of information on $\varphi $, that can obtained by a measuring
a quantum state which evolves in response to $\varphi $. The maximum QFI is related to the quantum Cramer-Rao lower bound (CRLB) for the minimum
variance any estimator \(\overset{\wedge }{\varphi }\), of $\varphi $ can possess, through:
\begin{equation}
\text{Var}\left[\overset{\wedge }{\varphi }\right]\geq \frac{1}{\text{QFI}_{\max }(\varphi )},
\end{equation}
where this inequality must be strict when the bound is derived assuming perfect, instantaneous measurements which can never be physically realised.
Often, we will use $\Delta \overset{\wedge }{\varphi } = \sqrt{\text{Var}\left[\overset{\wedge }{\varphi }\right]}$, as the minimum uncertainty
one can estimate $\varphi $, and refer to this quantity as sensitivity. We use several methods for calculating the QFI. First a non-constructive
method (15), which provides a bound for \(\text{QFI}_{\max }(\varphi )\) from the derivative of the Hamiltonian with respect
to $\varphi $, and assumes the spin is in a pure state
\begin{equation}
\text{QFI}_{\max }(\varphi )=\text{  }\left[\int _0^t\text{d$\tau $}\left.\left(\lambda _{\max }(\tau )-\lambda _{\min }(\tau )\right)\right/\hbar
\right]^2,
\end{equation}
where \(\lambda _{\max ,\min }\) are the maximum, minimum eigenvalues of \(\partial _{\varphi }H(\varphi )\). This method does not explicitly provide
the measurement that achieves the quantum CRLB, but can be used as an idealised benchmark to compare to specific measurements. To calculate \(\text{QFI}_{\max
}(\varphi )\) we differentiate Eq.\,(1) when the signal field is applied, to obtain

\begin{equation}
\partial _{\varphi }H(\varphi )= \frac{i \hbar  \Omega }{2} \left(
\begin{array}{cc}
 0 & -e^{-i \left(\varphi +t \omega _s\right)} \\
 e^{i \left(\varphi +t \omega _s\right)} & 0 \\
\end{array}
\right)
\end{equation}
which has eigenvalues \(\lambda =\pm \hbar  \Omega /2\). The maximum QFI and \(\Delta \overset{\wedge }{\varphi }\) for estimating \(\varphi\) is
then
\begin{equation}
\text{QFI}_{\max }(\varphi )=t^2\Omega ^2,        \Delta \hat{\varphi } \geq  \frac{1}{t \Omega }.
\end{equation}

Pang and Jordan (15) use the same analysis to calculate the maximum QFI and uncertainty for estimating $\omega $ as: 
\begin{equation}
\text{QFI}_{\max }(\omega )=t^4\Omega ^2, \;\;\; \Delta \hat{\varphi } \geq  \frac{1}{t^2 \Omega },
\end{equation}
from
\begin{equation}
\partial _{\omega }H(\omega )= \frac{i \hbar  \Omega  t}{2} \left(
\begin{array}{cc}
 0 & -e^{-i \left(\varphi +t \omega _s\right)} \\
 e^{i \left(\varphi +t \omega _s\right)} & 0 \\
\end{array}
\right).
\end{equation}
One can also calculate the classical Fisher information from the result of a given measurement (c.f. Eq.\,(14)):
\begin{equation}
\text{FI}(\varphi ) = \left(\frac{\partial \Pr (1;\varphi )}{\partial \varphi }\right)^2\frac{1}{\Pr (1;\varphi ) (1-\Pr (1;\varphi ))},
\end{equation}
and see whether the measurement achieves the maximum QFI. This method operates on an actual data set, but requires that an estimator exists that
satisfies the regularity condition and achieves the bound. If the Fisher information of a specific measurement attains the maximum QFI, then we can
say that the measurement is optimal. For a data-set containing more than one measurement we use the general form for calculating the minimal variance
obtained from an R-point dataset
\begin{equation}
\text{Var}\left[\overset{\wedge }{\varphi },R\right]\geq 1\left/ \underset{n=1}{\overset{R}{\sum }}\text{FI}(\varphi ,n)\right.
\end{equation}

Eq. (36) states that the overall FI from a complete dataset is the summation of the information that each \(n^{\text{th}}\) datapoint provides. In
practise we are assuming that a deterministic signal with an unknown parameter $\varphi $ is imprinted onto the sensor and readout of the sensor
provides a time-trace of data-points, forming an R-point data set $\{$x[1], x[2], ... x[R]$\}$, which must depend on $\varphi $ in order to estimate
it. The value of each data-point is inherently random due to noise (specifically quantum projection noise). By taking into account how the dataset 
depends on $\varphi $ and the noise in the data, we can obtain a bound on estimating $\varphi $. In the case that the probability and therefore the FI does not explicitly depend on n, i.e. it is the same for each datapoint, then the measurements are identical and independently distributed and we have 

\begin{equation}
\begin{array}{cc}
\text{Var}\left[\overset{\wedge }{\varphi },R\right]\geq  1\left/\underset{n=1}{\overset{R}{\sum }}\text{FI}(\varphi )\right.=\frac{1}{R \text{FI}(\varphi)}\\
\Delta \hat{\varphi } \geq  \frac{1}{\sqrt{R}\sqrt{\text{FI}(\varphi )}}.\\
\end{array}
\end{equation}

Performing R measurements on a single spin to estimate \(\overset{\wedge }{\varphi }\), we then have
\begin{equation}
\text{Var}\left[\overset{\wedge }{\varphi },R\right]\geq  \frac{1}{R t^2\Omega ^2},\; \; \; \Delta \hat{\varphi } \geq  \frac{1}{\sqrt{R}t \Omega }=\frac{1}{\sqrt{T
t} \Omega }
\end{equation}
where $T = Rt$ is the total measurement time required to obtain all the data. It is easy to check that Eq.\,(12) is a factor of $\pi $ above this limit.
For a dataset that depends on $\delta $, the relative detuning between signal and control fields can also be estimated. First we note that the relative
signal phase at the start of the \(n^{\text{th}}\) measurement is 

\begin{equation}
\varphi _n=\varphi _{n-1}+\tau \delta  = \varphi +(n-1)\tau  \delta ,
\end{equation}

for $n \in $ $\{$1, 2, 3, ... ,R$\}$, \(\varphi _1 \equiv \varphi \equiv \varphi \left(t_0\right)\) and where $\tau $ is the time between subsequent
measurements, which is the time spent performing unitary evolution plus time spent reading out the qubit. For a dataset obtained by repeatedly applying
\(U_{\text{seq}}\), in Eq.\,(6) to the atom for time $t$ and reading out the population, the detuning becomes imprinted onto the data by affecting the
phase at the start of each measurement. As a result we expect to obtain a dataset where Pr(1), oscillates in time with frequency
$n \tau \delta $. If we assume that the optimal way to measure the signal frequency is by estimating the signal phase at the
start of each experiment, then we can assert that the variance obtained from the following probability distribution is minimal
\begin{equation}
\Pr \left(1;t,\delta _s,\delta _c,\varphi ,n\right) \approx  \text{cos}\left[\frac{\varphi }{2}+\frac{n \tau  \delta }{4}\right]^2 \text{sin}[t \Omega /2]^2.
\end{equation}
Here we stress that the probability as written above is not exact. In particular, if the signal and control fields are detuned from each other and
from atomic resonance, then this formula is an approximation. The estimate of $\varphi $ should depend on the detuning more intrinsically
than simply updating the phase at each measurement, for example if the sensor is tuned completely out of resonance we should get nearly no information
about $\varphi $. We note that measurements sampled from such a probability distribution are optimal for estimating $\varphi $ at every time and cannot be surpassed, therefore any modifications
to this formula to make it more exact must come at a cost of reducing the ability to estimate $\varphi $. However, this formula is useful, since
it depends on $\delta $ and allows an analytic limit on estimating $\delta$ to be obtained. Assuming each measurement is instantaneous
then $\tau $ = t. Then the information and uncertainty obtained from the \(n^{\text{th}}\) measurement is

\begin{equation}
\begin{array}{cc}
\text{FI}(\delta ,n) = -\frac{n^2 t^2 \text{sin}\left[\frac{1}{4} (n t \delta +2 \varphi )\right]^2 \text{sin}\left[\frac{t \Omega }{2}\right]^2}{3-\text{cos}\left[\frac{n
t \delta }{2}+\varphi \right]+2 \text{cos}\left[\frac{1}{4} (n t \delta +2 \varphi )\right]^2 \text{cos}[t \Omega ]}\\
\\
\Delta \hat{\delta}_n \geq  \frac{\sqrt{3-\text{cos}\left[\frac{n t \delta }{2}+\varphi \right]+2 \text{cos}\left[\frac{1}{4} (n t \delta +2 \varphi )\right]^2 \text{cos}[t
\Omega ]}}{n t \text{sin}\left[\frac{1}{4} (n t \delta +2 \varphi )\right] \text{sin}\left[\frac{t \Omega }{2}\right]}\\
\end{array}
\end{equation}

Evaluated at \(t_{\pi }=\pi /\Omega\) we have

\begin{equation}
\text{FI}(\delta ,n) = \frac{n^2 \pi ^2}{4 \Omega ^2},  \; \; \;    \Delta \hat{\delta }_n \geq  \frac{2\Omega }{n \pi }
\end{equation}

For times shorter than $\pi $/$\Omega $, the FI is maximised for $\varphi $ = $\pi $ and a series expansion around \(t= 0\) gives

\begin{equation}
\text{FI}(\delta ,n)\approx  \frac{1}{16} n^2 \Omega ^2 t^4 + O\left[t^5\right]
\end{equation}

\subsection*{SI Note 5: Estimating the initial signal phase $\varphi $ using a NOON state}

To generate \(|\Psi _{\text{NOON}}\rangle =\frac{1}{\sqrt{2}}(|00\rangle +|11\rangle )\) from \(|\Psi _{00}\rangle =|00\rangle\) using direct
coupling between the atoms requires entangling gates. Entanglement generation involves the following unitaries: a $\pi $/2-rotation on qubit 1, to obtain the state \(|x\,0\rangle\), then
a C$_1$NOT$_2$ gate (i.e. $\pi $-rotation on qubit 2 controlled by qubit 1). If we restrict ourselves to preparing the sensing state with
control fields, then generation of \(|\Psi _{\text{NOON}}\rangle\) takes at least three times longer than generation of the state \(|xx\rangle\),
which simply requires a $\pi $/2-rotation performed simultaneously on both qubits.\\
Interaction with the signal for a time $\pi/(2 \Omega $), followed by two disentangling gates, brings the system to $|$11$\rangle $ with a probability

\begin{equation}
\Pr (11)=1-\left(\frac{1}{2}-\frac{1}{2} \text{cos}[2 \phi ]\right)^2.
\end{equation}

Neglecting the fact that the time to perform the C$_1$NOT$_2$ gate is limited by the coupling between the atoms, and by increasing their interaction they
become detuned to the signal, the experiment takes more than 3 times longer than performing two $\pi $/2-rotations (mediated by the signal and control
fields) on both qubits in parallel.

\subsection*{SI References}
1. Schmitt, S., Gefen, T., Louzon, D. et al. Optimal frequency measurements with quantum probes. npj Quantum Inf 7, 55 (2021). https://doi.org/10.1038/s41534-021-00391-5

\subsection*{Appendix}
Response from \emph{Nature} (submitted 07-04-2020):\\
``Thank you for submitting your manuscript "Frequency measurements beyond the Heisenberg time-energy limit with a single atom" to Nature. I regret to say, however, that we are unable to offer to publish it.

Owing to the fact that we receive more papers than we can publish, we decline a substantial proportion of manuscripts without sending them to referees, so that they may be sent elsewhere without delay. Decisions of this kind are made by the editorial staff when it appears that a paper, even if the referees were to certify its technical correctness, is unlikely to succeed in the competition for limited space. These editorial judgments are based on such considerations as the degree of advance provided, timeliness and the breadth of potential interest to the wider research community.

In the present case, we appreciate the specialist interest likely to be generated by your measurement of the frequency of a near-resonant field with a single atom with uncertainty below the Heisenberg limit. Indeed, your findings will no doubt be of value to those in a position to further build on your experiment as well as your theoretical proposal of a new fundamental uncertainty limit. I regret, however, that we are unable to conclude that the paper in itself provides the sort of leap in scientific understanding that would be likely to excite the immediate interest of the broad and diverse readership of Nature. We therefore feel that the present paper would find a more appropriate audience in a journal that publishes more specialised research.

I am sorry that we could not be more positive in this occasion, but we thank you for your interest in the journal. Please be assured that this editorial decision does not represent a criticism of the quality of your work, nor are we questioning its value to others working in this area. We hope that you will rapidly receive a more favourable response elsewhere."\\

Response from \emph{Science} (submitted 19-04-2020):\\
``Thank you for submitting your manuscript "Frequency measurements beyond the Heisenberg time-energy limit with a single atom" to Science. Because your manuscript was not given a high priority rating during the initial screening process, we have decided not to proceed to in-depth review. The overall view is that the scope and focus of your paper make it more appropriate for a more specialized journal. We are therefore notifying you so that you can seek publication elsewhere.

We now receive many more interesting papers than we can publish. We therefore send for in-depth review only those papers most likely to be ultimately published in Science. Papers are selected on the basis of discipline, novelty, and general significance, in addition to the usual criteria for publication in specialized journals. Therefore, our decision is not necessarily a reflection of the quality of your research but rather of our stringent space limitations."\\

Response from \emph{PNAS} (submitted 03-05-2020):\\
``Thank you for submitting your manuscript, titled "Frequency measurements beyond the Heisenberg time-energy limit with a single atom", to PNAS; the results of our assessment have led us to the decision to decline to consider it for publication at this time.

PNAS is a multidisciplinary journal that aims to publish high-impact research of general interest to the scientific community. Because we receive more than 18,000 submissions every year, incoming manuscripts undergo an initial evaluation by a member of the Editorial Board, who is also a member of the National Academy of Sciences, to determine whether the potential novelty, impact, and relevance in the broad scientific community merit further detailed technical review. In your case, our assessment is that your manuscript does not meet one or more of the principal aims of our journal and on this basis we expect that the likelihood that detailed review will lead to publication is low.

This decision is necessarily subjective and does not reflect an evaluation of the technical quality of your work or of its appropriateness for a more specialized audience; accordingly, we wish you success in finding a more suitable venue for publication soon."\\

Editor's Remarks to Author:
``I found reading this article to be extremely difficult. The basic idea or goal is not clearly explained, although it seems to describe a method to measure the phase of a ``signal" field relative to a ``control" field by applying these fields to generate pi/2 pulses to a single atom or two-level system, whose state changes are then measured. In any case, the basic idea/goal wasn't clear. The relation to the Heisenberg time-energy uncertainty principle is not clearly explained. The mathematical expressions are stated but not derived."

\end{document}